\documentclass[twocolumn,amsmath,amssymb,aps,showpacs]{revtex4}
\usepackage{epsfig,graphicx,amsmath,amsfonts,amssymb ,eucal}
\newcommand{\half}{\frac{1}{2}}
\newcommand{\ddn}[1]{\frac{\partial #1}{\partial n}}

\begin{document}
\title{Loop Dynamics in DNA Denaturation}
\author{A. Bar$^{1}$, Y. Kafri$^{2}$ and D. Mukamel$^{1}$}
\affiliation{$^1$Department of Physics of Complex Systems,
Weizmann Institute of Science, Rehovot, Israel 76100.\\
$^2$Department of Physics, Technion, Haifa, Israel 32000.}
\begin{abstract}
The dynamics of a loop in DNA molecules at the denaturation
transition is studied by scaling arguments and numerical
simulations. The autocorrelation function of the state of
complementary bases (either closed or open) is calculated. The
long-time decay of the autocorrelation function is expressed in
terms of the loop exponent $c$ both for homopolymers and
heteropolymers. This suggests an experimental method for measuring
the exponent $c$ using florescence correlation spectroscopy.
\end{abstract}

\maketitle

The thermodynamic properties of DNA near the thermal denaturation
transition have been extensively studied during the last few
decades \cite{Kafri-2002,review1}. At low temperatures a small
fraction of the base pairs are unbound, forming loops of
fluctuating lengths. These loops increase in size as the
temperature is raised, until the denaturation transition is
reached and the two strands separate. Experiments using uv
absorption and specific heat measurements have yielded valuable
information on equilibrium properties of DNA \cite{review1}.
Recently, single molecules techniques, most notably Fluorescence
Correlation Spectroscopy (FCS) have been used to study dynamical
properties such as the temporal behavior of loops
\cite{altan-bonnet}.

The main theoretical approach for studying DNA denaturation has
been introduced by Poland and Scheraga (PS) \cite{ps} and was used
to analyze the case of homopolymers. It was found that the
dependence of the entropy of a loop on its length plays a dominant
role in determining the thermodynamic behavior near the
transition. On general grounds one can argue that the entropy of a
loop of length $n$ takes the form $S=k_B \log(\Omega(n))$, where
$\Omega(n)\sim s^n/{n^c}$ is the number of loop configurations.
Here $s$ is a model-dependent constant and $c$ is a universal
exponent whose numerical value has been debated over the years and
was found to depend on the degree in which excluded volume
interactions are taken into account \cite{ps,Fisher,Kafri-2000}.
When excluded volume interactions both within a loop and between
the loop and the rest of the chain are taken into account one
finds $c \simeq 2.12$ \cite{Kafri-2000,Kafri-2002}. This result,
which predicts a first order denaturation transition, has been
verified numerically \cite{carlon}. While numerical studies of the
model with excluded volume interaction yield a clear first order
transition \cite{Causo-2000}, a direct experimental measurement of
$c$ is rather difficult and has not been carried out so far.
Theoretical studies of the case of a heteropolymers suggest that
disorder makes the transition of order higher than two
\cite{ChateTang,Toninelli}.

In this paper we analyze the loop dynamics at the denaturation
transition. The analysis suggests a method for measuring the
exponent $c$. We focus on predictions for FCS studies
\cite{altan-bonnet}. In these experiments one monitors the state
of a base pair (whether it is open or closed) as a function of
time. The measured quantity is the base pair autocorrelation
function $C_i(t)=\langle u_i(0)u_i(t) \rangle$ where $u_i(t)=1,0$
is a variable which indicates if base pair $i$ is open $(1)$ or
closed $(0)$ at time $t$. By analyzing the loop dynamics using a
scaling approach and by direct modelling we express the temporal
behavior of the autocorrelation function at the transition
temperature in terms of the exponent $c$.

Previous analyses of the loop dynamics have concentrated mainly on
the off critical region \cite{hanke,kats}. In these analyses the
loop is assumed to be in thermal equilibrium throughout its
evolution. As discussed below in detail, the validity of this
assumption is not obvious. In this work we test this crucial
assumption and demonstrate that it is valid at the denaturation
transition.

To proceed we consider the dynamics of a single isolated loop. In
this approach one ignores processes like merging of loops and the
splitting of a large loop into two or more smaller ones. This may
be justified by the fact that the cooperativity parameter, which
controls the statistical weight of opening a new loop, is
estimated to be rather small, $\sigma_0 \approx 10^{-4}$
\cite{carlon}. Thus splitting a loop into two is unfavorable.
Also, the average distance between loops, which within the PS
model is proportional to $1/\sigma_0$, is large, making the
independent loop approximation plausible. A loop may change its
size by closing or opening of base pairs at its two ends. It
survives as long as its two ends do not meet. Let $G(n,t)$ be the
survival probability of a loop of initial length $n$ for time $t$.
The equilibrium autocorrelation function, measured in FCS
experiments, is given by
\begin{equation}
C(t) \approx \frac{\sum_{n=1}^{\infty}P_{eq}(n)n G(n,t)
}{\sum_{n=1}^{\infty}P_{eq}(n)n} \label{eqn:corr}
\end{equation}
where for simplicity of notation we have dropped the site index
$i$. Here $P_{eq}(n)$ is the probability of having a loop of
length $n$ in equilibrium. The additional $n$ factor accounts for
the fact that site $i$ may be in any of the $n$ sites of the
initial loop. Note that we assume that site $i$ remains open as
long as the loop survives. This assumption does not affect the
behavior of the autocorrelation function in the scaling limit. An
interesting configuration is created when one end of the loop is
forced to be on a particular site. In this case no $n$ factor is
needed in Eq. \ref{eqn:corr} and the autocorrelation function for
the end of the loop reads
\begin{equation}
C_E(t) \approx \sum_{n=1}^{\infty}P_{eq}(n) G(n,t). \label{CE}
\end{equation}
Experimentally, this autocorrelation function may be measured by
studying a molecule which is clamped at one end with a mismatch
near this end. The autocorrelation function near the mismatch site
yields $C_E(t)$.

In the following we analyze the cases of homogeneous and
heterogenous DNA. We show that in the homogeneous case the
autocorrelation decays at large $t$ as $C(t)\sim t^{1-c/2}$ for
$c>2$ while it remains finite, $C(t)=1$, for $c<2$. On the other
hand we find $C_E(t)\sim t^{(1-c)/2}$ for $c>1$. Our analysis of
heteropolymers suggests that the disorder average of the
autocorrelation function behaves as $\overline{C}_E(t)\sim (\ln
t)^{2-2c}$ for $1<c<3/2$ and as $\overline {C}_E\sim (\log
t)^{-1}$ for $c>3/2$. Here the overline denotes an average over
disorder.

Consider first the case of a homopolymer. In this case it has been
shown that $P_{eq}(n)\sim n^{-c}e^{-n/\xi}$. The correlation
length $\xi$ diverges at the transition yielding $P_{eq}(n)\sim
n^{-c}$. In order to estimate the survival probability of a loop
of length $n$ we consider the dynamics of a loop under the
assumptions discussed above, where loops are non-interacting and
they do not split into a number of smaller loops. Similar to
\cite{hanke,kats} we further assume that the loop is in a local
thermal equilibrium at any given time during its evolution. The
validity of this assumption will be discussed in detail below. The
loop free energy is thus given by $f \propto n/\xi + c \ln n$
where $n$ is the loop size. Within the framework of the
Fokker-Planck equation, the probability distribution of finding a
loop of size $n$ at time $t$, $P(n,t)$, is given by
\begin{equation}
\frac{dP(n,t)}{dt} = D \ddn{} \left[\frac{1}{\xi} + \frac{c}{n} +
\ddn{}\right]P(n,t) \;,
        \label{eqn:FPE}
\end{equation}
where $D$ is the diffusion constant in base pair units. Here we
have taken the continuum limit and assumed the dynamics to be
over-damped. This equation has to be solved with the boundary
condition $P(0,t)=0$ and initial condition $P(n,0)=\delta(n-n_0)$.
The survival probability of the loop is given by
$G(n_0,t)=\int_{0}^{\infty}dnP(n,t)$. Using standard techniques
\cite{redner} it can be shown that at the transition temperature
($\xi^{-1}=0$) the survival probability obeys the scaling form
$G(n_0,t) = g\left(Dt/n_0^z\right)$ with $z=2$. The asymptotic
behavior of the scaling function for small and large values of the
argument is
\begin{equation}
g(x)\sim 1 \;\; {\rm for} \; x \ll 1  \;\; ; \;\; g(x)\sim
x^{-\frac{1+c}{2}} \;\; {\rm for} \; x \gg 1 \;.
\label{eq:asympsurvivh}
\end{equation}
Using these properties it is easy to calculate the long-time
behavior of the autocorrelation function (Eq. \ref{eqn:corr})
\begin{eqnarray}
C(t) \sim \left\{%
\begin{array}{ll}
    1 & \hbox{for}\;\;\; c\leq 2 \\
    t^{1-c/2} & \hbox{for}\;\;\; c > 2 \;.
\end{array}%
\right. \label{eq:autocorrhomo}
\end{eqnarray}
Thus, the asymptotic behavior of $C(t)$ could in principle be used
to measure the exponent $c$. In particular it can be used to
distinguish between the case of a continuous transition ($c \leq
2$), where $C(t)=1$, and a first order phase transition ($c>2$),
where $C(t)$ decays to zero. Similar analysis for the edge
autocorrelation function leads to $C_E(t)\sim t^{(1-c)/2}$ for
$c>1$.

A central assumption in the above analysis is that the loop is at
local equilibrium at any given time. A priori this is not
necessarily a valid assumption. The typical time for the survival
of a loop of length $n$ scales as $n^2$. On the other hand the
relaxation time of a loop configuration is also expected to scale
as $n^2$ when hydrodynamic interactions are ignored (to be
discussed below). Thus it is not clear that during the evolution
of the loop it is in local equilibrium. Away from the transition
point the loop size changes linearly in time and therefore the
assumption of local equilibrium is clearly not valid. In the
following we introduce a simple model for studying the loop
dynamics where hydrodynamic interactions are ignored. We find
strong evidence that the local equilibrium assumption holds
asymptotically even in this case.

\begin{figure}[h]
\centering
\includegraphics[scale=0.9]{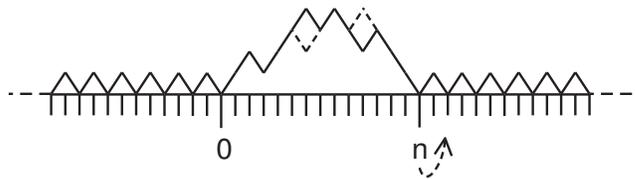}
\caption{A typical microscopic configuration of the loop in the
RSOS model. Dashed lines indicate possible dynamical moves of the
interface.} \label{fig:model}
\end{figure}

To this end we introduce and analyze a simple model for the loop
dynamics. This dynamics is described by a fluctuating interface
interacting with an attractive substrate in $d=1+1$ dimensions.
Here the interface height variable corresponds to the distance
between complementary bases. The interface configurations are
those of a restricted solid on solid (RSOS) model defined as
follows (see Fig. \ref{fig:model}): Let $h_i=0,1,2 \ldots$ be the
height of the interface at site $i$. The heights satisfy
$|h_i-h_{i+1}|= \pm 1$ and $h_i \geq 0$. Consider a loop between
sites $0$ and $n$ (where $n$ is even) as shown in Fig.
\ref{fig:model}. Outside the loop the interface is bound to the
substrate so that $h_{-2k}=h_{n+2k}=0$ for $k=0,1,\ldots$ while
for $0 \leq k \leq n$ the height $h_i$ can take any value which is
consistent with the RSOS conditions and is non-negative. We
consider a dynamics in which the loop is free to fluctuate and one
of its ends ($n \ge 0$) is free to move with the following rules.
For sites $2 \leq i \leq n-1$ the height is updates according to
\begin{equation}
h_i \to h_i \pm 2
\label{eq:looprate}
\end{equation}
with rate $1$ as long as $h_i \geq 0$ and the RSOS condition is
satisfied. For $i=n$ the loop length is changed according to the
following rules
\begin{eqnarray}
n &\to& n+2 \;\;\; {\rm with \; rate} \;\;\; \overline{\alpha}/4
\nonumber \\
n &\to& n-2 \;\;\; {\rm with \; rate} \;\;\; \overline{\alpha} \;,
\label{eq:endrate}
\end{eqnarray}
where $n$ can decrease only if $h_{n-2}=0$.

In principle one should let both ends fluctuate. However, for
simplicity, we consider the case where one of the ends is fixed.
It is straightforward to verify that the number of configurations
of a loop of size $n$ is given by $2^n/n^c$ with $c=3/2$ for large
$n$. Thus this model corresponds to a particular value of $c$.
However we expect similar results to hold for other values of $c$
as well. The ratio between the two length changing processes is
chosen such that in the large $n$ limit the loop is not biased to
either increase or decrease. This corresponds to the model being
at the denaturation transition point. The parameter
$\overline{\alpha}$ determines the rate of the length changing
processes: $\overline{\alpha}=0$ corresponds to the dynamics of a
loop of fixed length and as $\overline{\alpha}$ is increased the
length changing processes become faster. In a realization of this
dynamics at any given step one of $n$ possible moves is chosen. Of
these, $n-2$ moves correspond to an attempted update of the height
at sites $2,3, \ldots, n-1$. The other two moves correspond to an
attempt to update the edge by a move either to the right or to the
left. One attempted move of the edge defines a Monte Carlo sweep.
On average this amounts to updating all sites every two sweeps.
The numerical studies described below are done using $\overline
{\alpha} = 1$.

\begin{figure}[h]
\centering
\includegraphics[scale=0.75]{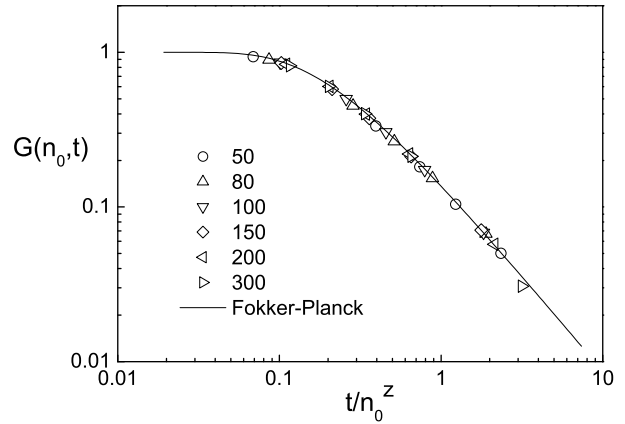}
\caption{ Data collapse of the survival probability (averaged over
$4 \cdot 10^4$ realizations) for some values of $n_0$ with
dynamical exponent $z=2.2$. The line corresponds a numerical
solution of Eq. \ref{eqn:FPE} } \label{CollapseZ2p2}
\end{figure}

In order to test the validity of the Fokker-Planck equation
(\ref{eqn:FPE}) for describing the dynamics of a loop we simulated
the dynamics of the model and calculated the survival probability
of a loop of initial length $n_0$. To this end an initial
configuration of a loop of fixed length $n_0$ is generated with
the correct equilibrium weight. Starting from this initial
configuration the dynamics is carried out. The results are
summarized in Fig. \ref{CollapseZ2p2} where the survival
probability is plotted as a function of the scaling variable
$t/n_0^z$ for several values of the loop size $n_0$. A very good
agreement with the predicted survival time obtained from the
solution of the Fokker-Planck equation (\ref{eqn:FPE}) is found.
However the optimal data collapse takes place at $z \approx 2.2$
rather than $z=2$. If this value of $z$ remains valid in the limit
of large $n$ it would imply that the local equilibrium assumption
is not valid.

In the following we argue that the value $z=2.2$ is due to finite
size effects and we expect that for large systems the value $z=2$
is recovered. To this end we calculate numerically the variance of
the loop size
\begin{equation}
w^2(t)=\langle (n(t)-\langle n(t) \rangle )^2 \rangle \;,
\label{eq:variancedef}
\end{equation}
where $\langle \cdot \rangle$ denotes an average over realizations
of the dynamics. We show that asymptotically it grows linearly
with time. This result indicates that the dynamical exponent is in
fact $z=2$ and that the deviations we observe for small $n_0$ are
due to finite size effects. We proceed by defining a variable
$\sigma_+(t)$ which takes the value $1$ if the length of the loop
increases at time $t$ and $0$ otherwise. Similarly, we define
$\sigma_-(t)$ and $\sigma_0(t)$ for steps which decrease the loop
size and steps in which the loop size does not change,
respectively. Clearly $\sigma_+(t)+\sigma_-(t)+\sigma_0(t)=1$. The
dynamics of the chain (\ref{eq:endrate}), implies that in the
limit of large $n_0$ one has
\begin{equation}
\langle \sigma_+(t) \rangle=\langle \sigma_-(t) \rangle = \alpha/8
\;\; ; \;\; \langle \sigma_0(t) \rangle=1-\alpha/4 \;,
\label{eq:moveprob}
\end{equation}
where $\alpha=\overline{\alpha}/ \max \{ 1, \overline{\alpha} \}$.
Denoting $U(t) \equiv \sigma_+(t) - \sigma_-(t)$, it is easy to
see that
\begin{eqnarray}
\frac{\Delta w^2(t)}{\Delta t} & \equiv & w^2(t)-w^2(t-1)
\nonumber \\
&=& 4\langle U(t)^2 \rangle +8 \sum_{\tau=1}^{t-1} \langle U(\tau)
U(t) \rangle \;, \label{eq:defvariancediff}
\end{eqnarray}
where
\begin{eqnarray}
 \langle U(\tau) U(t) \rangle &=& \langle \sigma_+(\tau) \sigma_+(t) \rangle+\langle
\sigma_-(\tau)
\sigma_-(t) \rangle \nonumber \\
&-&\langle \sigma_-(\tau) \sigma_+(t) \rangle -\langle
\sigma_+(\tau) \sigma_-(t) \rangle \;. \label{eq:sigmaSrelation}
\end{eqnarray}

\begin{figure}[h]
\centering
\includegraphics[scale=0.75]{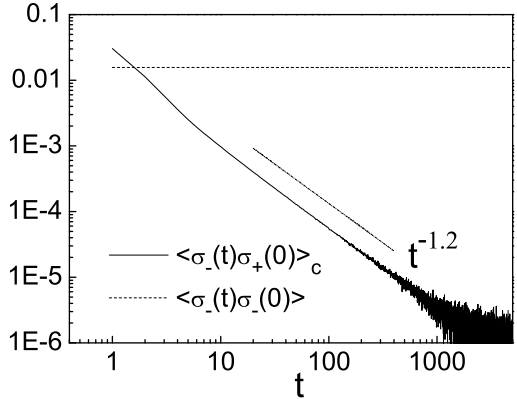}
\caption{Correlation functions of the $\sigma$ variables as
obtained by averaging over $1.9 \cdot 10^5$ realizations.}
\label{Correlations}
\end{figure}

It is evident that a loop increasing step at time $t$,
($\sigma_+(t)=1$), is uncorrelated with steps which took place at
time $\tau<t$. Thus $\langle \sigma_+(\tau) \sigma_+(t) \rangle =
\langle \sigma_-(\tau) \sigma_+(t) \rangle = \alpha^2/64$.
Numerically we find $\langle \sigma_-(\tau) \sigma_-(t)
\rangle=\alpha^2/64$ (see Fig. \ref{Correlations}). Using these
result we finally obtain
\begin{equation}
\frac{\Delta w^2(t)}{\Delta t}=  \alpha - 8 \sum_{\tau=1}^{t-1}
\left[ \langle \sigma_+(\tau) \sigma_-(t) \rangle_c \right] \;.
\label{eq:varfinal}
\end{equation}
with $ \langle \sigma_+(\tau) \sigma_-(t) \rangle_c \equiv \langle
\sigma_+(\tau) \sigma_-(t) \rangle - \frac{\alpha^2}{64}$.
Numerical simulations of the dynamics show strong correlation
between $\sigma_+(\tau)$ and $\sigma_-(t)$ with an algebraic decay
in $t-\tau$ (see Fig. \ref{Correlations}). It is interesting to
note that the dynamics of the chain induces such long range
temporal correlations mediated by the loop dynamics.

\begin{figure}[h]
\centering
\includegraphics[scale=0.75]{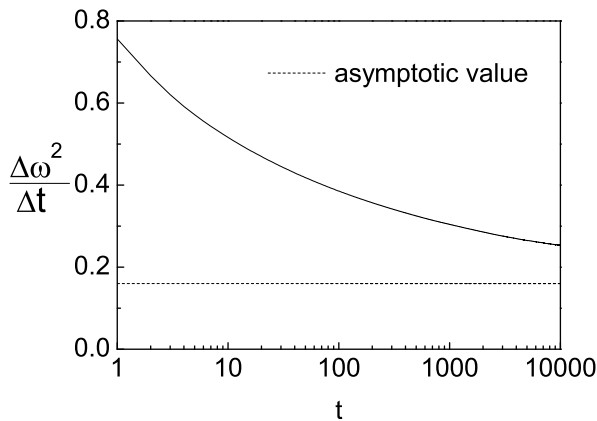}
\caption{ Temporal behavior of $\Delta w^2/ \Delta t$ showing its
slow convergence to its asymptotic non-vanishing value.}
\label{dw2dt}
\end{figure}

By extrapolating the sum on the right hand side of Eq.
\ref{eq:varfinal} using the asymptotic form $At^{-\gamma}$ with
$A\approx 0.015$ and $\gamma\approx 1.2$, deduced from Fig.
\ref{Correlations}, we find that the sum converges to a value
$\approx 0.84<\alpha=1$ indicating that $w^2(t)\approx 0.16t$ at
large $t$, which in turn yields $z=2$ (see Fig.\ref{dw2dt}). The
slow power-law convergence towards the asymptotic value implies
that it may require large systems to observe the long time
behavior of the autocorrelation function, (\ref{eq:autocorrhomo}).

We now turn to the heteropolymer case. To avoid the typical
denominator problem in disordered systems we only consider the
autocorrelation function $C_E(t)$ and study its disorder average.
For a heteropolymer the binding energy of the $i$th base-pair,
$\eta(i)$, (and therefore also the length changing rate
$\overline{\alpha}_i$) are taken to be uncorrelated quenched
random variables. As the edge moves $m$ steps the binding energy
changes by $\Delta(m)=\sum_{i=1}^m \eta(i)$. The rate
$\overline{\alpha}_i$ is simply related to the binding energy
$\eta(i)$, similar to the dynamics of DNA unzipping
\cite{lubensky}. Since the variance of $\Delta(m)$ grows linearly
with $m$ we can safely neglect the effect of logarithmic
correction to the loop entropy on the dynamics. At the transition
point the dynamics of the loop length corresponds to that of an
unbiased walker on a {\it random forcing} energy landscapes. It is
known that the probability of a walker (representing the edge of
the loop) not to return to the origin on such a disordered energy
landscape, $G_d(n,t)$, has the scaling form
$G_d(n,t)=g_d\left((\log t)^2/n_0\right)$ \cite{ledoussal}. The
asymptotic behavior of the scaling function for small and large
values of the argument is
\begin{equation}
g_d(x)\sim 1 \;\; {\rm for} \; x \ll 1  \;\; ; \;\; g_d(x)\sim
x^{-\half} \;\; {\rm for} \; x \gg 1 \;.
\label{eq:scalingsinaiasmpt}
\end{equation}
This result is universal and independent of the potential
realization. Note that in this case there is a separation of time
scales where the typical survival time of the loop is much longer
than the loop relaxation time. Thus the use of local equilibrium
dynamics is clearly justified.

To complete this analysis one has to evaluate the equilibrium loop
size distribution. Extensive numerical studies suggest that the
disorder average loop statistics remain of the same form with the
same exponent $c$ as in the case of homopolymers \cite{carlon,
garel}.

Combining this with the universal form of the survival probability
we finally reach the asymptotic form of disordered average
autocorrelation function:
\begin{eqnarray}
\overline{C}_E(t) \sim \left\{%
\begin{array}{ll}
    (\log t)^{2-2c} & \hbox{for}\;\;\; 1< c \leq 3/2\\
    (\log t)^{-1} & \hbox{for}\;\;\; c > 3/2 \\
\end{array}%
\right.
\label{eq:autocorrhetero}
\end{eqnarray}

We conclude with a comment on hydrodynamic interactions. In the
present study these interactions have not been included. It is
well known that relaxation processes of polymers in solutions are
faster when hydrodynamic interactions are taken into account.
Within the Zimm model \cite{Doi} it scales with the length of the
polymer as $n^{3 \nu}$. Recent experiments on single-stranded DNA
have measures a scaling $n^{3/2}$ \cite{Oleg}. This fast
relaxation time lends further support for the local equilibrium
hypothesis applied in this work.

\noindent This work is supported by the Minerva Foundation with
funding from the Federal Ministry for Education and Research, by
the Albert Einstein Minerva Center for Theoretical Physics and by
the US-Israel Binational Science Foundation (BSF). We thank the
Newton Institute in Cambridge (UK) for the kind hospitality during
the programme "Principles of the Dynamics of Non-Equilibrium
Systems" where part of this work was carried out.

\end{document}